\documentclass[12pt]{article}
\usepackage{amssymb} \usepackage{amsfonts}
\usepackage{graphicx}
\begin{document}
\title{\bf{Neutron star structure in a Quark Model with Excluded Volume Correction}
\thanks{\it{This work was partially supported by the CONICET, Argentina.}}}
\author{R. M. Aguirre and A. L. De Paoli.\\
Departamento de F\'{\i}sica, Fac. de Ciencias Exactas,\\
Universidad Nacional de La Plata.\\
C. C. 67 (1900) La Plata, Argentina.}
\maketitle
\begin{abstract}

We study the effects of the finite size of baryons on the equation
of state of homogeneous hadronic matter. The finite extension of
hadrons is introduced in order to improve the performance of field
theoretical models at very high densities. We simulate the
in-medium averaged baryon-baryon strong repulsion at very short
distances by introducing a Van der Waals like normalization of the
fields. This is done in the framework of the Quark Meson Coupling
model that allows to take care of the quark structure of baryons.
Since within this model the confinement volume evolves with the
fields configuration, the treatment is not equivalent to a simple
hard-core potential. We investigate the phase transition to quark
matter and the structure of neutron stars. We have found
significant corrections at high densities.\\


\end{abstract}
\newpage

\renewcommand{\theequation}{\arabic{section}.\arabic{equation}}

\section{Introduction}

Investigation of hadronic matter at extreme conditions of density
and temperature is a current issue of research, since the study of
its properties will eventually shed some light over the recovering
of QCD symmetries \cite{HATSUDA,BROWN}. The phase diagram of
hadronic matter is expected to be very complex, exhibiting exotic
phases like superfluidity, meson condensates, dibaryon condensate,
etc. The gradual emergence of quark droplets would finally lead to
a transition to deconfined quark matter. Every one of these
phenomena affect the equation of state (EOS) and could have
macroscopic manifestations as, for example, in the structure of
stars.

In most hadronic matter studies baryons are assumed to be
point-like. This can be justified because at densities below the
nuclear matter saturation density the finite volume effects are
expected to be small. However, the spatial extension of baryons
was recognized as an essential point in the study of the
collective phenomena at very high densities \cite{KAPUSTA}.

It is worthy to mention that there are only a few field
theoretical models which consistently include the baryonic spatial
extension. The most commonly used are the Skyrme and bag-like
models. In the first case the inclusion of finite baryon density
effects is not straightforward due to the topological character of
its solutions \cite{RAKHIMOV}. On the other hand, further
refinements of the original MIT bag model allow to deal with
medium effects upon the hadron structure
\cite{Gui,ST,SR,PAL,PANDA}. Within this scheme there were recent
efforts to include the repulsion between overlapping bags, by
means of effective
 short-ranged quark-quark correlations \cite{SR}.

The authors attempted to take into account finite volume
correction previously, using a Van der Waals-like method to study
nuclear matter with lambda hyperons \cite{OUR}. The total volume
appearing in the thermodynamical quantities was replaced by the
available volume, in accordance with related investigations
\cite{Waa1,Waa2,StoGre,Cley,SINGH}.

In the present work we generalize this approach to study the
properties of hadronic matter including the octet of low lying
baryons, and to check out their influence over the transition to
quark matter. The resulting EOS is applied to study the
composition of neutron stars. Following our previous study
\cite{OUR}, we introduce these corrections at the level of the
normalization of the baryon fields in the Quark Meson Coupling
(QMC) model \cite{Gui,ST}. The motivation for such a procedure is
to parameterize in compact form the strong baryon-baryon repulsion
at very short distances. We focus on the high density regime, thus
it is justified to consider the statistical average of the
interaction instead of looking at its details.

 In the QMC model the size of the confining
volume has its own dynamical evolution which takes into account
the baryonic density and the fields configuration, and it is
obtained in a self-consistent calculation. Thus it can somehow be
regarded as an effective degree of freedom. The bag radius changes
smoothly with the medium properties, and in this sense the finite
volume corrections introduced here cannot be considered as a hard
core interaction.

In the next section we give a resume of the QMC model and we
introduce the excluded volume correlations. In section 3 we
describe neutron star matter and the phase transition to quark
matter. Numerical results and discussions are given in section 4,
and conclusions are drawn in section 5.

\section{The Quark Meson Coupling Model} \setcounter{equation}{0}

 Relativistic Hadron Field Theories provide a good description of
nuclear matter near the saturation density, and of finite nuclei
too. For this purpose only a small number of free parameters is
required. Mean field approximation (MFA) is suited for the aim of
these theories, since the treatment of matter at medium and high
densities do not require the detailed structure of the
interactions. Within this framework the QMC model \cite{Gui,ST}
can be viewed as an extension of the Quantum Hadro-Dynamics (QHD)
models \cite{Wa,SW}.

In the QMC model baryons are represented as non-overlapping
spherical bags containing three valence quarks; the bag radius
changes dynamically with the medium density. Baryons interact by
the exchange of $\sigma$, $\omega$ and $\rho$ mesons coupled
directly to the confined quarks. It has been found that these
extra degrees of freedom, provided by the internal structure of
the baryon, lead to quite acceptable values of the nuclear matter
compressibility at saturation. Despite of the explicit quark
fields in the QMC model, hadronic thermodynamical properties are
evaluated in such a way that baryons are handled as point-like
particles with an effective mass $M_B^{\ast}$ which depends on the
$\sigma$ field.

In the MFA the Dirac equation for a quark of flavor $q, \; (q=u,
d, s)$, of current mass $m_q$ and $I_{3}^q$ third isospin
component, is given by

\begin{equation}
( i \gamma^{\mu} \partial_{\mu} - g_{\omega}^q \gamma^0 \omega_0 -
g_{\rho}^q I_{3}^q\; \gamma^0 b_0 - {m_q}^\ast) \Psi^q = 0.
\label{QMCEQ}
\end{equation}
In this equation all meson fields have been replaced by their mean
field values. Mesons  couple linearly only to non-strange quarks,
i.e. $g_{\sigma}^s=g_{\omega}^s=g_{\rho}^s=0$. Therefore the
parameters ${m_q}^{\ast}$  are given by
\begin{eqnarray}
{m_{u,d}}^\ast&=&m_{u,d} - g_{\sigma}^{u,d} \sigma, \nonumber \\
{m_s}^{\ast}&=&m_s. \label{QMASS}
\end{eqnarray}
For a spherically symmetric bag of radius $R_b$ representing a
baryon of class $b$, the normalized quark wave function
$\Psi^q_b(r,t)$ is given by

\begin{equation}
\Psi^q_b(r,t)={\cal N}_b^{-1/2} \frac{e^{-i{\varepsilon}_{q b} t}}
{\sqrt{4\pi}} \left( \begin{array}{c}
j_0 (x_{qb} \, r/R_b) \\
i \beta_{q b} {\vec{\sigma}}.{\hat{r}} j_1 (x_{q b} \, r/R_b)
\end{array} \right) \chi ^q,
\end{equation}
where $\chi ^q $ is the quark spinor and

\begin{equation}
\varepsilon_{q b} = \frac{\Omega_{q b}}{\!R_b} + g_{\omega}^q
\;\omega_0 + g_{\rho}^q I_{3}^q\; b_0,
\end{equation}

\begin{equation}
{\cal N }_b={R_b}^3\;[2 \Omega_{q b} (\Omega_{q b} - 1) + R_b
{m_q}^\ast ]\; \frac{ j_0^2 (x_{q b}) }{x_{q b}^2},
\end{equation}

\begin{equation}
\beta_{q b} ={\left[ \frac{\Omega_{q b} - R_b {m_q}^\ast
}{\Omega_{q b} + R_b {m_q}^\ast } \right]}^{1/2},
\end{equation}
with $\Omega_{q b} =[x_{q b}^2 +{(R_b {m_q}^\ast)}^2    ]^{1/2}$.
The eigenvalue $x_{q b}$ is solution of the equation

\begin{equation}
j_0 (x_{q b}) = \beta_q \; j_1 (x_{q b}), \label{BOUNDARY}
\end{equation}
which arises from the boundary condition at the bag surface.\\ In
this model the ground state bag energy is identified with the
baryon mass $M_b^\ast$,

\begin{equation}
M_b^\ast=\frac{\sum_q n_q^b \Omega_{q b} - z_{0 b}}{R_b} +
\frac{4}{3} \pi B   {R_b}^3, \label{BAGMASS}
\end{equation}
where $n_q^b$ is the number of quarks of flavor $q$ inside the
bag. The bag constant $B$ represents the difference of energy per
unit volume between the vacuum with and without broken QCD
symmetry. It is numerically adjusted to get definite values for
the proton bag radius. The zero-point motion parameters $z_{0 b}$
are fixed to reproduce the baryon spectrum at zero density.\\

Eq. (\ref{BAGMASS}) shows that the baryon effective mass is a
function of the bag radius $R_b$. In the original MIT bag
calculations $R_b$ is a constant fixed at zero baryonic density,
but in the QMC it is a variable dynamically adjusted to reach the
equilibrium of the bag in the dense hadronic medium. We use the
equilibrium condition proposed in ref. \cite{OUR}, which results
by imposing a vanishing net flux of the energy-momentum tensor
through the surface of the bag immersed in the dense hadronic
medium. This gives

\begin{equation}
- \frac {1} {4 \pi R_b^2} {\left( \frac {\partial{M_b}^\ast}
{\partial R_b} \right)}_{\sigma, x_{q b}} =
 \frac{1}{3{\pi}^2 \, \xi} \sum_{b'} \int_0^{k_{b'}}
\frac{dk k^4} {\sqrt{{{M_{b'}}^\ast}^2+k^2}}, \label{QMCc}
\end{equation}
where $\xi=1$. The interested reader can find the detailed
derivation of this relation in \cite{OUR}.

This result reflects the balance of the internal pressure of the
bag with the baryonic contribution to the total external pressure,
represented by the left and right sides of Eq.(\ref{QMCc}),
respectively.

The factor $\xi$ will be redefined below when excluded volume
effects will be considered.

 It must be noted that Eq.(\ref{QMCc}) differs from the standard QMC
condition \cite{ST}, i.e.
\[ {\left( \frac {\partial {M_{b}}^{\ast}} {\partial R_b}
\right)}_{\sigma}=0. \] Both prescriptions coincide only in the
case of vanishing density. Eq.(\ref{QMCc})could be more
appropriate for dealing with finite density calculations,
nevertheless there remains to elucidate the problem of overlapping
bags as the density grows. Therefore it becomes necessary to
introduce some further considerations into the formalism to handle
this feature.

Once ${M_{b}}^\ast$ has been defined microscopically, the hadronic
thermodynamics in the QMC model resembles that of the Quantum
Hadrodynamics. In the MFA for homogeneous infinite static matter
all meson fields are replaced by their averaged values, i. e.

\begin{eqnarray}
\sigma&=&\;\;\;\sigma_0\;\;\;\;\,=-\frac{1}{{m_\sigma}^2} \sum_b
\frac{d{M_b}^\ast}{d\sigma} n_s^b, \label{SIGMA} \\
\omega_{\mu}&=&\;\omega_0
\delta_{\mu0}\;\;\,=\;\;\frac{1}{{m_\omega}^2} \sum_B g_{\omega}^b
n^b \; \delta_{\mu0}, \label{OMEGA}\\ b_{\mu}^a&=&b_0
\delta_{\mu0} \delta_{a3}=\;\;\frac{1}{{m_\rho}^2} \sum_b
g_{\rho}^b I_{3}^b n^b \; \delta_{\mu0} \delta_{a3}, \label{RHO}
\end{eqnarray}
where $a=1, 2, 3$ runs over all isospin directions and $I_{3}^b$
is the third isospin component of baryon $b$. In our calculations
we have used the values $m_\sigma=550 MeV$, $m_\omega=783 MeV$, and
$m_\rho=770 MeV$ for the meson masses.\\

The dispersion relation for the b-baryon is
\begin{equation}
k_0^b=\sqrt{{{M_b}^\ast}^2+{\vec{k}}^2} \pm g_{\omega}^b \omega_0
\pm g_{\rho}^b I_{3}^b b_0, \label{PARTENER}
\end{equation}
for particle $(+)$ and antiparticle $(-)$ solutions. Within the
MFA at zero temperature only the particle solutions contribute.\\

The scalar ($n_s^b$) and baryonic  ($n^b$) densities are defined
with respect to the ground state of the hadronic matter $|GS>$
composed of baryons filling the Fermi sea up to the state with
momentum $k_b$
\begin{eqnarray}
\!n_s^b \!\!\!&=& \!\!\!\!\!<GS| {\bar{\Psi}}^b \;\Psi^b
|GS>=\vartheta\;\frac{1} {{\pi}^2} {M_ b}^\ast \int_0^{k_b} dk
\frac{k^2} { \sqrt{{M_b^\ast}^2+k^2}}, \label{SCALDENS} \\ \!n^b
\!\!\!&=& \!\!\!\!\!<GS| {{\Psi}^{\dag}}^b {\Psi}^b |GS> =
\;\vartheta\; \frac {{k_b}^3} {3{\pi}^2}. \label{VECDENS}
\end{eqnarray}
In Eqs.  (\ref{SCALDENS}) and (\ref{VECDENS} ) the factor
$\vartheta$ is included for future use and it takes the value
$\vartheta = 1$ for point-like baryons.

In the next section we describe hadronic matter in
$\beta$-equilibrium and electrically neutral, therefore we also
consider leptons treated as free Dirac particles. The leptonic
density $n_l$ is related to the Fermi momentum $k_l$ by
\begin{equation}
n^l = <GS| {{\Psi}^{\dag}}^l \Psi^l |GS> = \frac {{k_l}^3}
{3{\pi}^2}. \label{LEPNUM}
\end{equation}
Given a distribution of baryonic species we can calculate the
total energy density $\epsilon_H$ and pressure $P_0$ of hadronic
matter for point-like baryons
\begin{eqnarray}
\epsilon_H&=&\frac {1}{2} {m_\sigma}^2\; {\sigma_0}^2 +\frac
{1}{2} {m_\omega}^2\; {\omega_0}^2 +\frac {1}{2} {m_\rho}^2\;
{b_0}^2 \nonumber \\ &+& \frac{\vartheta}{{\pi}^2} \sum_b
\int_0^{k_b} dk k^2 \sqrt{{M_b^\ast}^2+k^2} \nonumber \\
&+&\frac{1}{{\pi}^2} \sum_l \int_0^{k_l} dk k^2
\sqrt{{m_l}^2+k^2}, \label{ENERGY} \\ P_0&=&\sum_b \mu_0^b n^b +
\sum_l \mu^l n^l - \epsilon_H, \label{PRESSURE}
\end{eqnarray}
\noindent
 where $\mu_0^b=k_0^b (k_b)$ (see Eq. (\ref{PARTENER}))
and $\mu^l$ are the chemical potentials for point-like baryons and
leptons, respectively.

The QMC model has been widely used to describe the dense hadronic
matter, it is based in a great extent on the assumption of
non-overlapping bags. Therefore the breakdown of this hypotheses
signals the limit of applicability of the model. Using this
criterion an upper density limit around three times the saturation
density of symmetric nuclear matter has been found \cite{SR}. For
densities beyond this value the quark-quark interactions through
the confinement region should be introduced, so the naive bag
picture is not sufficient to describe the physical situation in
this case. This fact is taken into account in the literature using
different approaches. For example in \cite{SR} the quark-quark
correlations between bags are introduced in the overlapping
region. A different approach is given in \cite{PAL}, where the
effects of quark-quark correlations are assumed to be represented
by the exponential dependence of the bag constant $B$ on the
$\sigma$ meson.

 We propose an alternative viewpoint that preserves the
scheme of non-overlapping bags, and that intends to take care of
the strong repulsive component of the baryon-baryon interaction
that appears as a consequence of the internal structure of the
particles. This short range repulsion can be treated in a
simplified model where baryons are described using a dynamics of
extended objects. Therefore the fraction of available space is
reduced as compared to the case of point-like particles. A similar
approach has been applied to study the phase transition of nuclear
matter to the quark-gluon plasma \cite{Waa1,Waa2} and in heavy-ion
collisions \cite{StoGre,Cley}.

Since finite size baryons are assumed as non-overlapping, their
motion must be restricted to the available space $V'$ defined as
\cite{Waa1,Waa2}
\begin{equation}
V'=V-\sum_b N^b v_b,
\end{equation}
with $N^b$ the total number of baryons of class $b$ inside the
volume $V$, and $v_b$ the effective volume per baryon of this
class. Hence we conjecture that one can renormalize the particle
(antiparticle) wave function replacing $V'$ for $V$, and thus the
effective baryon fields $\Psi$ can be written as
\begin{eqnarray}
{\Psi}^b(x)={(V')}^{-1/2} \sum_{\vec{k},s} &[& a^{b}(\vec{k},s)
u^b(\vec{k},s) e^{-ik^{\mu} x_{\mu} } \nonumber \\
&+&b^{b \dag}(\vec{k},s) v^b(\vec{k},s) e^{\;ik^{\mu} x_{\mu} }
\;]\label{RENFIELD}
\end{eqnarray}
in terms of the Fock space operators $a$ and $b$, for particle and
antiparticle respectively. In this way the finite size of the
baryons is automatically accounted for into the field dynamics.

Eq. (\ref{RENFIELD}) reinforces the fact that, within this model,
the inner and outer regions of the bag must be regarded as
complementary in order to avoid inconsistencies. On the contrary,
if these regions can overlap the validity of the linear (Eq.
(\ref{BOUNDARY})) and non-linear (Eq. (\ref{QMCc})) boundary
conditions should be revised and additional degrees of freedom,
such as exotic multi-quark states, should be included. Therefore
our assumption extends to higher densities the quasi-particle
picture and it generates further nonlinear baryonic couplings.

For the moment the volumes $v_b$ are parameters associated with
the trial quantum state of the whole system, and will be
determined using a variational principle, as it is explained
later.

It is interesting to note that for a mixture of different baryons
the excluded volume is not exactly the same for all the species
\cite{Waa3}. To simplify the discussions, in this paper we neglect
these small differences.

The effective volume per baryon $v_b$ is proportional to the
actual baryon volume, i.e. for spherical volumes of radius $R_b$
\begin{equation}
v_b=\alpha \frac{4 \pi}{3} {R_b}^3,
\end{equation}
and for sharp rigid spheres $\alpha$ is a real number ranging from
$4$, in the low density limit, to $3\sqrt{2}/{\pi}$, which
corresponds to the maximum density allowed for non overlapping
spheres, in a face centered cubic arrange. Since we wish to study
the high density regime of homogeneous isotropic matter, we shall
adopt $\alpha= 3\sqrt{2}/{\pi}$ in all our calculations. Thus
$v_b=4\sqrt{2}{R_b}^3$, and the limit of validity of the
calculations would correspond to densities $n_{max}$ such that the
center of mass of baryons are at a distance greater than $2R$
apart. This gives $n_{max}=\sqrt{2}/(8{R_{max}}^3)$, where
$R_{max}$ denotes the biggest radius among all present baryonic
classes. As we shall see below, with the implementation of the
procedure outlined in Eq. (\ref{RENFIELD}) this limit is never
reached in the range of densities explored in the present
calculations.\\

In order to see how excluded volume corrections appear in our
approach, we shall use the renormalized field of Eq.
(\ref{RENFIELD})  to calculate the relationship among the baryonic
densities and the Fermi momenta $k_b$

\begin{equation}
\label{DENSREN} n^b = V^{-1} \int_V dx^3 <GS| {{\Psi}^{\dag}}^b
{\Psi}^b |GS>= (1-\sum_{b'} n^{b'} v_{b'})\;\frac {{k_b}^3}
{3{\pi}^2},
\end{equation}
where $\sum\limits_{\vec{k}} \rightarrow V'/(2\pi^3) \int dk^3$
has been used.\\ This result is equivalent to Eq. (\ref{VECDENS})
if the factor $\vartheta$ takes the value
\begin{equation}
\vartheta=1-\sum_b n^b  v_b,  \label{THETA}
\end{equation}
for finite baryonic effective volumes $v_b$ . In the limit $v_b
\rightarrow 0 $ one recovers the point-like expressions.

Eq. (\ref{DENSREN}) shows that these kind of short-range
correlations couple non linearly the baryons among themselves, in
a density dependent way. Thus baryons are considered in this
scheme as quasi-particles dressed with these corrections.

The density $n_b$ for a given baryonic species appears on both
sides of Eq.(\ref{DENSREN}), and it is possible to solve it
exactly for $n_b$ in terms of all the Fermi momenta , namely
\begin{equation}
n^b=\frac{1}{(1+\sum\limits_{b'}\frac{{k_{b'}}^3}{3{\pi}^2}\;
v_{b'})} \;\frac{{k_b}^3}{3{\pi}^2}. \label{BARNUM}
\end{equation}
Since $\vartheta$ depends explicitly upon the
baryonic densities,  the chemical potentials get an extra term,
i.e.
\begin{eqnarray}
\mu^b&=&{ \left( \frac{\partial\epsilon_H}{\partial n_b} \right)
}_ {{ n_{{b'}_{{}_{{}_{{}_{\!\!\!\!\!\!\!\!\!\!\!b' \neq b}}}}}}}
= \mu_0^b + \Delta \mu^b, \label{PQ} \\ \Delta \mu^b &=&\frac{v_b
}{3{\pi}^2} \sum_{b'} \int_0^{k_{b'}} \frac{dk k^4}
{\sqrt{{{M_{b'}}^\ast}^2+k^2}} \label{POTQUIM2}
\end{eqnarray}
\noindent
 where the energy density $\epsilon_H$ is given by
Eq.(\ref{ENERGY}) with $\vartheta$ defined as in Eq.(\ref{THETA}).

The total pressure acquires an additional term $\Delta P$ as
compared to the pressure of point-like baryons $P_0$ in Eq.
(\ref{PRESSURE})
\begin{eqnarray}
P_H=P_0 + \Delta P = P_0 + \sum\limits_{b}  n^b \Delta \mu^b.
\label{TRUEPRESS}
\end{eqnarray}

We proceed to give a physical interpretation to the volumes $v_b$
of the baryons, or equivalently their radii $R_b$. They are
considered within this approach as variational parameters of the
trial quantum state of the system. Their equilibrium values, at
zero temperature, must be determined by minimizing the total
energy of the whole system, in agreement with \cite{Waa1,Waa2}.
This procedure defines unambiguously the equilibrium radii of the
bags, that depend on the baryonic density.\\

The new equilibrium condition for the bag consistent with the
excluded volume corrections introduced in Eq. (\ref{RENFIELD}) has
the same form as that given in Eq.(\ref{QMCc}), but now $\xi=
{n^b_s}/{(\alpha \, n^b)}$.

As we shall see in the next sections, as the density grows a state
of deconfined quarks becomes more favorable than a system of
quarks confined within baryons. Because of this, the concept of
excluded volume is meaningless when the baryonic phase solves
completely.\\

In our approach only the baryonic states receive an explicit
correction due to short range correlations, by the normalization
of the fields $\Psi^b$. It must be stressed that the meson mean
field values are completely determined from the baryonic sources,
that already include finite size corrections in $\vartheta$.
Leptons do not couple to strong interactions, and in this sense
they are taken as point-like particles.

To summarize, in the QMC model extended to high densities the
hadronic matter properties are determined applying the set of
equations (\ref{BOUNDARY}) to (\ref{ENERGY}), together with
(\ref{PQ})-(\ref{TRUEPRESS}) for a fixed total baryonic density
$n$, using the value of $\vartheta$ given in (\ref{THETA}).

\section{Quark matter phase transition and the structure of neutron stars}
\setcounter{equation}{0}

Neutron star matter is electrically neutral and it has reached
equilibrium against $\beta$-decay.  The relative abundance of the
different baryonic species are determined by these conditions. The
structure of neutron stars with hyperon contributions has been
widely studied, for recent investigations see for example
\cite{PAL,STARS} and references therein. In particular in
\cite{PAL} a extension of the QMC model is used, that takes into
account the density dependence of the bag constant $B$ and
includes strange mesons.

In the present work we consider the nucleon duplet (n,p), the
$\Lambda$-hyperon, the $\Sigma$ triplet, the $\Xi$ duplet, and two
lepton species, electron and  muon. \\

In hadronic matter at chemical equilibrium the following relationships
are fulfilled for the baryonic and leptonic chemical potentials
\begin{eqnarray}
\mu^b&=&\mu^n+ Q_b \mu^e \;\;\;\; {\mathrm {(if \; baryon\;
class\;  b\;
is \; present)}}, \label{CHEMEQ}\\ 
 \mu^e&=&\mu^{\mu} \;\;\;\;\;\;\;\;\;\;\;\;\;\;\;\; {\mathrm {(if \; muons \; are \;
present)}},
\end{eqnarray}
with $Q_b$ the electric charge in units of the positron charge for
the class $b$ of baryons, $\mu^b$ given by Eq. (\ref{PQ}) and
$\mu^l = k_0^l(k_l) =\sqrt{{m_l}^2+{k_l}^2}$.\\

Mean field equations are solved for fixed total baryonic density
$n$, and zero total electric charge density
\begin{eqnarray}
n&=&\sum_b n^b, \label{baryonum}\\
0&=&\sum_b Q^b n^b- \sum_l n^l,  \label{CHARGE}
\end{eqnarray}
where the sums run over the baryonic octet and over the two lepton
species, respectively.

Once Eqs. (\ref{BOUNDARY})-(\ref{QMCc}) have been solved, from
Eqs. (\ref{SIGMA})-(\ref{RHO}) we determine the meson fields
($\sigma_0 , \omega_0, b_0$) and together with Eqs.
(\ref{VECDENS}),(\ref{CHEMEQ})-(\ref{CHARGE}) we get the baryonic
and leptonic densities ($n^b, n^l$), all the Fermi momenta ($k_b,
k_l$) and chemical potentials ($\mu^b, \mu^l$).\\

As the baryonic density increases a phase transition from hadronic
to quark matter, made up with deconfined quarks, can be reached.
Therefore the previous set of equations must be modified to
satisfy the new conditions of $\beta$-equilibrium in the
transition region. Since the baryonic number and the electric
charge neutrality are always conserved, a smooth crossover between
hadronic and quark matter must be expected because there are two
conserved charges. In fact, in this coexistence region of mixed
hadron and quark phases the total baryonic density and the total
(zero) electric charge are shared between this two phases.\\

To determine the composition of the transition region we suppose
that quarks can exist either in the confined phase (inside baryons)
or as deconfined particles. As a first approximation we suppose that
the deconfined phase contains free quarks and leptons, and
non-perturbative gluon effects are represented by the bag constant $B$.

Under equilibrium conditions, the net flux of quarks through the bag
surface is zero, and hence the two phases coexist with the only
constraint of globally conserved electric and baryonic charges.

We adopt the treatment of \cite{GLEND2}, as it is appropriate to
describe phase transitions with more than one conserved charge.
Thus, in the coexistence phase, the conservation equations
(\ref{baryonum}) and (\ref{CHARGE}) are generalized to
\cite{GLEND2}

\begin{eqnarray}
n&=&(1-\chi)n_H + \chi n_Q, \nonumber \\
&& \label{MIXEDEQ}\\
 0&=&(1-\chi)\sum_b Q^b n^b+ \chi \sum_q Q^q n^q - \sum_l  n^l,
\nonumber
\end{eqnarray}
where $n_H=\sum_b n^b$, $n_Q=\sum_q n^q/3$ are respectively the
hadron and quark contribution to the baryon number density;
$n^q=N_c k_q^3/(3 \pi^2)$ is the number density of quarks for
$N_c=3$ colors, $k_q$ is the Fermi momentum , and $Q^q$ is the
electric charge in units of the positron charge for quarks of
flavor $q$. The quantity $\chi$ is the volume fraction
corresponding to the quark matter phase $(0 \le \chi \le 1)$. The
$\beta$-equilibrium condition for quarks in the deconfined phase
reads: $\mu_d=\mu_s=\mu_u+\mu_e$, with $\mu_q=\sqrt{m_q^2+k_q^2}$.
In the mixed phase these relations must be supplemented with
$\mu_n=3 \mu_d-\mu_e$ and the mechanical equilibrium condition
$P_H=P_Q$, where $P_{H,Q}$ are the pressures in each phase. It
must be noted that the case $\chi=0 (1)$ in Eq. (\ref{MIXEDEQ})
corresponds to the pure hadronic (quark) matter phase instance,
the mixed phase is in between.

The energy density in the mixed phase can be similarly expressed
as $\epsilon=(1-\chi)\epsilon_H+ \chi \epsilon_Q$. The quantities
$\epsilon_Q$ and $P_Q$ are, respectively, the energy density and
pressure for the deconfined phase which contains
free quarks and leptons \cite{GLEND2}

\begin{eqnarray}
\epsilon_Q&=&B+ \frac{N_c}{\pi^2} \; \sum_q \int_0^{k_q} dk k^2
\sqrt{{m_q}^2+k^2} \nonumber \\
& & \;\;\;\; + \; \frac{1}{{\pi}^2} \; \sum_l \int_0^{k_l} dk k^2
\sqrt{{m_l}^2+k^2},
\label{QENERGY} \\
P_Q&=&\sum_q \mu_q n^q + \sum_l \mu_l n^l - \epsilon_Q.
\label{QPRESSURE}
\end{eqnarray}

Non-perturbative effects in $\epsilon_Q$ arising
from the gluons are represented by the bag constant $B$ \cite{GLEND2}.

Of course, when the pure quark matter phase is reached ($\chi=1$),
the conditions of chemical equilibrium apply among deconfined
quarks, since this means that baryons have solved completely.

The EOS emerging from this calculation can be used to evaluate the
properties of neutron stars. The stellar radius $R$ and mass $M$
are obtained by solving the Tolman-Oppenheimer-Volkoff
relativistic equations for a spherically symmetric (non-rotating)
neutron star
\begin{eqnarray}
\frac{dP}{dr}&=&- (G/c^2) \frac{[\epsilon(r)+P(r)]\,[m(r)+4 \pi
r^3 P(r)/c^2]}{r^2[1-2 (G/c^2) m(r)/r]} \, , \nonumber \\
m(r)&=&\int_0^r 4 \pi \, {r'}^2 \, [\epsilon(r')/c^2] \, dr' \, .
\label{MASSR}
\end{eqnarray}
Starting from a given value $\epsilon_c$ for the central energy
density, these equations are integrated outward until a radius
$R(\epsilon_c)$ is reached for which the pressure $P$ is zero, and
$M=m(R)$ is defined. \\

To determine more accurately the radius of the star, it becomes
necessary to use an appropriate EOS for low densities. We have
selected the EOS given in reference \cite{BPS} for baryonic
densities below $0.1 n_0$. It is worthful to mention that another
possibility is to choose the EOS given in \cite{NEGVAU}, which
proves to be very similar to the former \cite{BPS}.

The moment of inertia $I$ for a slowly rotating star can be
obtained using \cite{BPS,AKMAL}
\begin{equation}
I=\frac{8 \pi}{3}\int_0^{R} r^4 \, e^{-\nu (r)/2} \,
\frac{[\epsilon(r)+P(r)]/c^2}{\sqrt{1-2 (G/c^2) m(r)/r}}\;
\frac{\bar{\omega}(r)}{\Omega} \, dr \, ,
\label{INERR}
\end{equation}
where $\Omega$ is the uniform angular velocity of the star as seen
by a distant inertial observer, $\Omega \ll (c/R) \, \sqrt{(G/c^2)
M/R}$. The radial function $\nu (r)$ is solution of the
differential equation
\begin{equation}
\frac{d \nu}{dr}=2 \, (G/c^2) \, \frac{[m(r)+ 4 \pi r^3
P(r)/c^2]}{r^2 [1-2 (G/c^2) m(r)/r ]} \, ,
\end{equation}
with the boundary condition $\nu (R)=- \ln[1-2 (G/c^2) M/R]$. The
relative angular velocity $\bar{\omega}(r)$ measured with respect
to the local dragged inertial frame, is the solution of
\begin{equation}
\frac{d}{dr}\left[ r^4 j(r) \, \frac{d \bar{\omega}(r)}{dr}\right]
+ 4 r^3 \, \bar{\omega}(r) \, \frac{dj}{dr} = 0 \, ,
\end{equation}
with $j(r)=\sqrt{1-2 (G/c^2) m(r)/r} \exp[-\nu (r)/2]$ inside the
star.  Since $j(r)~=~1$ for $r \geq R$, $\bar{\omega}(r)$ has the
form $\bar{\omega}(r)=\Omega [1-2 (G/c^2) \, I/r^3]$ outside.

A useful quantity in astronomical observations is the surface
redshift $z$ given by
\begin{equation}
z={[\,1-2 (G/c^2) M/R]\,}^{-1/2} - 1 \, .
\end{equation}

Numerical values obtained in our approach can be found in Table
\ref{TABLEII}.

\section{Numerical Results} \setcounter{equation}{0}

 Within the present model the quark masses
take the current values $m_u=m_d=5 MeV$ and $m_s=150 MeV$. The
parameter $B$  is the same for all the baryon bags, and we avoid
any speculation about its density dependence. For a given value of
$B$ the set of parameters $z_{0b}$ are adjusted to obtain the
experimental baryon masses at zero density.

Since mesons interact directly with quarks, the corresponding
meson-baryon couplings are related to the quark-meson couplings
$g_{\phi}^q$ ($\phi = \sigma, \omega, \rho $, $q= u,d$) in a
simple way \cite{ST}. Denoting as $g_\phi^b$ the coupling of the
$\phi$-meson to $b$-baryon,
\begin{eqnarray}
g_{\sigma}^{b} &=& N^b_{ns} g_{\sigma}^{u}, \nonumber \\
g_{\omega}^{b} &=& N^b_{ns} g_{\omega}^{u},  \\ g_{\rho}^{b} &=&
g_{\rho}^{u}, \nonumber
\end{eqnarray}
where $N^b_{ns}$ is the non-strange quark number inside the baryon
$b$. Thus, for given $g_{\sigma}^{u}, g_{\omega}^{u}$ and
$g_{\rho}^{u}$, the full set of baryon-meson couplings can be
determined. Their numerical values are obtained by reproducing the
symmetric nuclear matter properties at saturation, i.e. baryonic
density, binding energy and symmetry energy
\begin{eqnarray}
& & n_0 = \;0.15 fm^{-3},  \nonumber \\ & & E_b= {(\epsilon/n)}_0
-M c^2 = -16 MeV,  \nonumber \\ & & a_s = \frac{1}{2} {\left(
\frac{\partial^2 (\epsilon/n)}{{\partial t}^2} \right)}_{t=0}
=\;35 MeV,
\end{eqnarray}
where $t=(n_n - n_p)/n$ and $M = 938.92 MeV/{c^2}$ is the averaged
free nucleon rest mass.

As expected, the values of the couplings are sensitive to either the
inclusion or not of the excluded volume corrections. Both instances are
considered in Table \ref{TABLEI}.

For practical applications, we have chosen the values
$B^{1/4}=169.93, 187.83$ and $210.85\, MeV$ that will be denoted
respectively as (a), (b) and (c) in the following. They yield a
proton bag radius $R_p=0.8, 0.7$ and $0.6$ fm respectively (see
Table \ref{TABLEI}). The cases with the excluded volume
corrections (CC) have been compared with calculations without them
(NC). We remark that in the cases labeled NC we take $\xi =
\vartheta=1$, and $\Delta \mu^b=0$ for all baryons.

The same $B$ is used consistently in the QMC and deconfined quark
descriptions.

In all cases we start in the hadronic phase increasing the
baryonic density until the coexistence conditions with quark
matter are satisfied. The exception here is the case (c)-NC that
stays always in the hadronic phase, at least within the range of
validity of the non-overlapping bags assumption.\\ This phase
transition occurs at critical densities $n_{cl}$ which are shown
in Table \ref{TABLEI}; it can be seen that $n_{cl}$ increases with
$B$. The lowest value $n_{cl}\simeq n_0$ corresponds to
$B^{1/4}=169.93 MeV$, and therefore we interpret that it is not
suitable for a true physical consideration and has been kept only
for illustrative purposes.

For the bag parameters $B$ considered here, the non-overlapping
bag hypotheses breaks down in the NC approach at limiting
densities $n_{max}$, before the pure quark plasma state has taken
place. These upper limits are displayed in the last column of
Table \ref{TABLEI}. Thus, stars with a pure quark core could not
be described within this framework.

On the other hand, in the CC instance the upper density threshold
$n_{cu}$ for the mixed phase is enlarged as $B$ increases. No bag
overlapping is produced in the calculations using the CC option
for densities $n<n_{cu}$, and for $n>n_{cu}$ excluded volume
effects are meaningless because only quark matter without hadrons
can be found.

\begin{figure}[ht]
\centering
\includegraphics[width=0.8\textwidth]{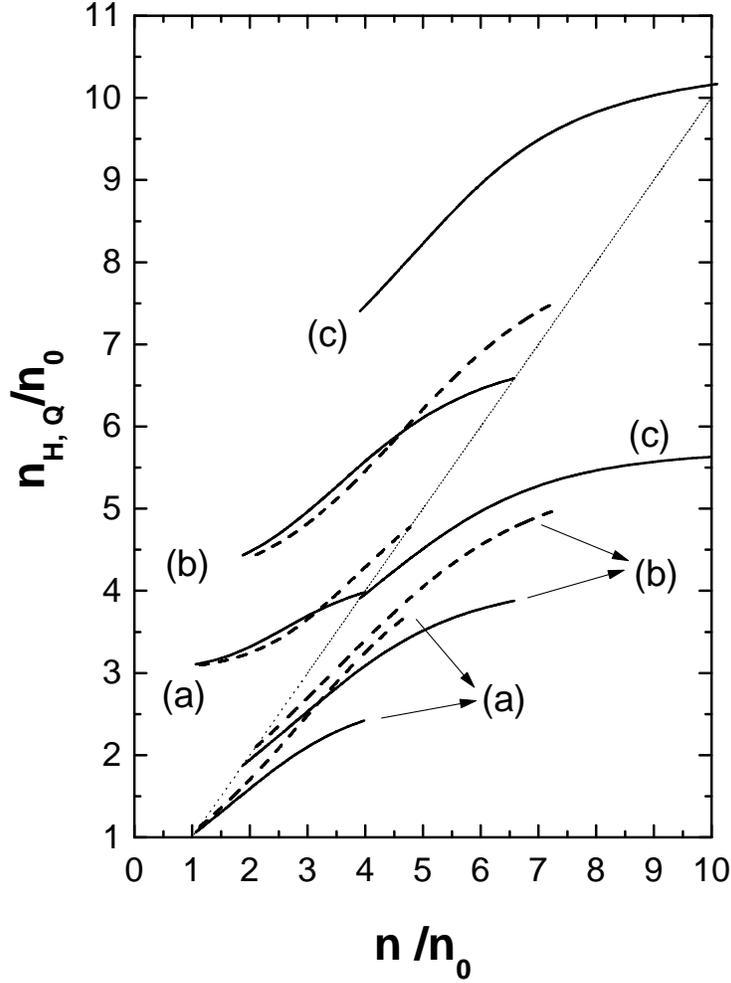}
 \caption{\footnotesize{Hadronic ($n_H$) and quark ($n_Q$)
contributions to the total baryonic density in the mixed phase.
The results with (without) excluded volume correction are
represented with solid (dashed) lines. The different bag constant
values $B^{1/4}=169.9, 187.8$, and $210.8$ MeV are distinguished
with the labels $a, b$ and $c$ respectively. For comparison we
have also drawn a curve corresponding to a pure phase (dotted
line). All curves above (below) this line represent the
contributions $n_Q$ ($n_H$). The case (c)-NC stays in the pure
hadronic phase for all the range of densities studied, and
therefore coincides with the dotted line in this graph. The CC
curves are plotted up to the upper density threshold $n_{cu}$ .
Numerical values for the density threshold $n_{c l}$ and $n_{cu}$
can be seen in Table~\ref{TABLEI}}.}\label{FIG1}
\end{figure}

In Fig. \ref{FIG1} we plot the contribution of the hadronic phase
($n_H$) and the free quark phase ($n_Q$) to the total baryonic
density in the mixed phase, see upper line of Eq. (\ref{MIXEDEQ}).
For the sake of completeness we draw a dotted line corresponding
to matter in a pure (either hadronic or quark) phase, all curves
above (below) this line represent the contributions $n_Q$ ($n_H$).
The case (c)-NC stays in the pure hadronic phase for all the range
of densities studied, and therefore it coincides with the dotted
line in this plot. It can be seen that $n_H$ lies always below 5.6
$n_0$, which corresponds to the upmost case ((c)-CC)) considered
here.

\begin{figure}[ht]
\centering
\includegraphics[width=0.95\textwidth]{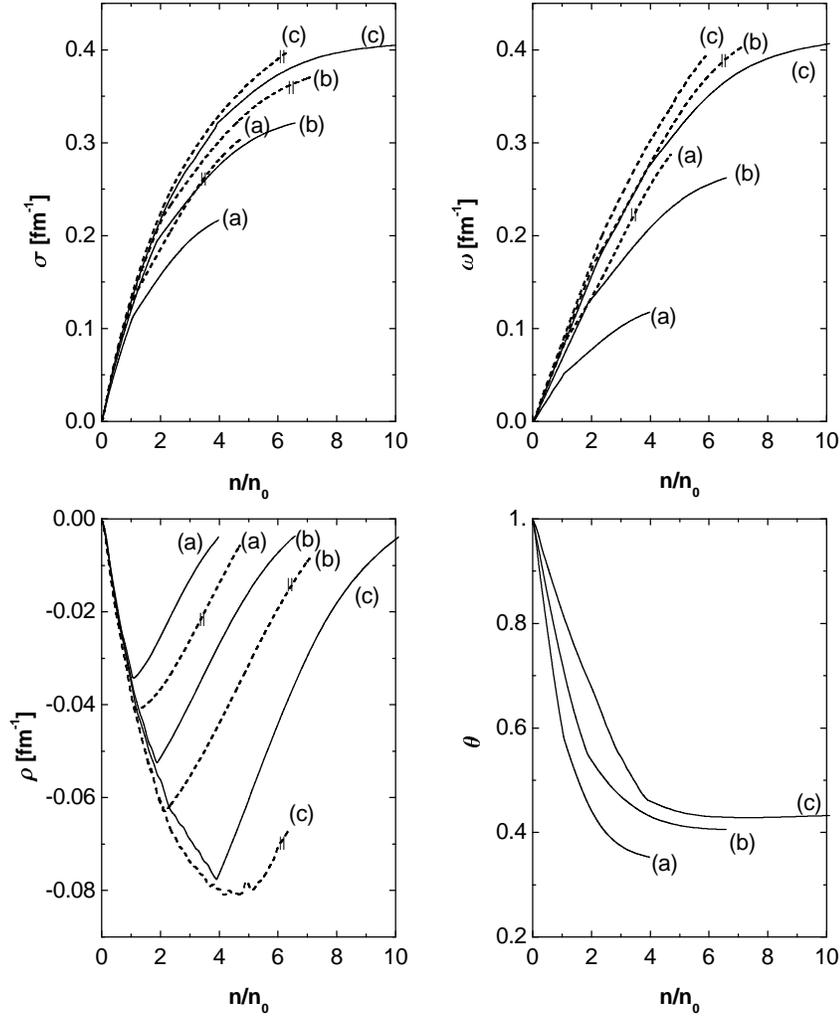}
 \caption{\footnotesize{Meson mean field values
$\sigma, \omega$ and $b_0$ as functions of the baryonic density
relative to the saturation nuclear density $n_0$. The line and
label conventions are the same as in Fig. \ref{FIG1}. In the right
lower corner the volume correction factor $\vartheta$ is displayed
in terms of the baryonic density. The double vertical bars
indicate the limit of the non overlap assumption for the NC
cases.} }\label{FIG2}
\end{figure}

\begin{figure}[ht]
\centering
\includegraphics[width=0.95\textwidth]{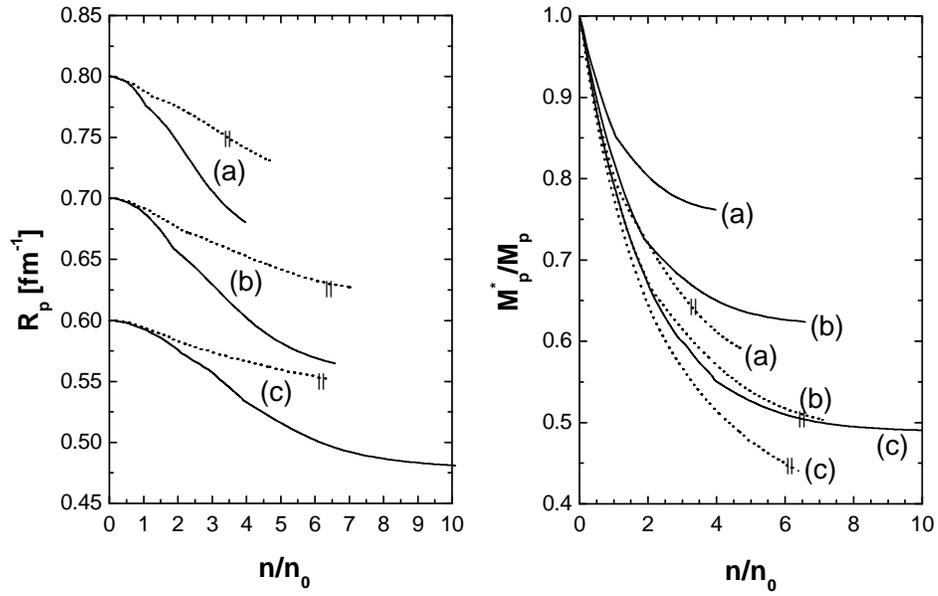}
 \caption{\footnotesize{The in-medium proton bag
radius $R_p$ and proton mass $M_p^\ast$ relative to its vacuum
values. The line and label convention are the same as in Fig.
\ref{FIG2}.}}\label{FIG3}
\end{figure}

In Fig. \ref{FIG2} we show the meson mean field values in the
hadronic and mixed phases. Differences between the CC and NC
treatments become appreciable for $n>n_0$. As can be seen, the
$\rho$ meson amplitude has a sudden change of slope at the phase
transition due to the change in the isospin composition of the
hadronic sector in the neighborhood of $n_{cl}$. In the same
figure the factor $\vartheta$ containing the excluded volume
correction is displayed as a function of the baryonic density.
This factor seems to become almost constant at
sufficiently high densities.\\
 The proton bag radius $R_p$ and its effective mass
$M_p^\ast$ can be examined in Fig. \ref{FIG3}. A faster radius
decrease is observed in the CC instance, as compared with the
respective NC calculation. This fact explains the absence of
overlap and the consequent lengthening of the validity range of
the CC approach.  A dropping of about $15\%-20\%$ in $R_p$ is
predicted before the hadronic matter has completely disappeared;
this strong compression of the baryon bags helps the deconfinement
mechanism in the CC case. This can be understood because at a
given density the pressure and the chemical potential of baryons
are slightly greater when the excluded volume correlations are
included, than in the NC results. Bearing in mind that at the
transition point there is a crossing of the curves representing
the pressure in terms of the chemical potential for the hadronic
and quark phases, this crossing occurs at lower densities for the
CC than for the NC case. On the other side, due to the slowly
decrease of the bag radius in the NC approach, the overlapping of
bags can happen before the phase transition to pure quark matter
takes place.

 The effective mass $M_p^\ast$ exhibits a
monotonous decrease as a function of the baryonic density, the
rate of variation at low densities is attenuated by both
decreasing $B$ and/or including volume corrections.

\begin{figure}[ht]
\centering
\includegraphics[width=0.95\textwidth]{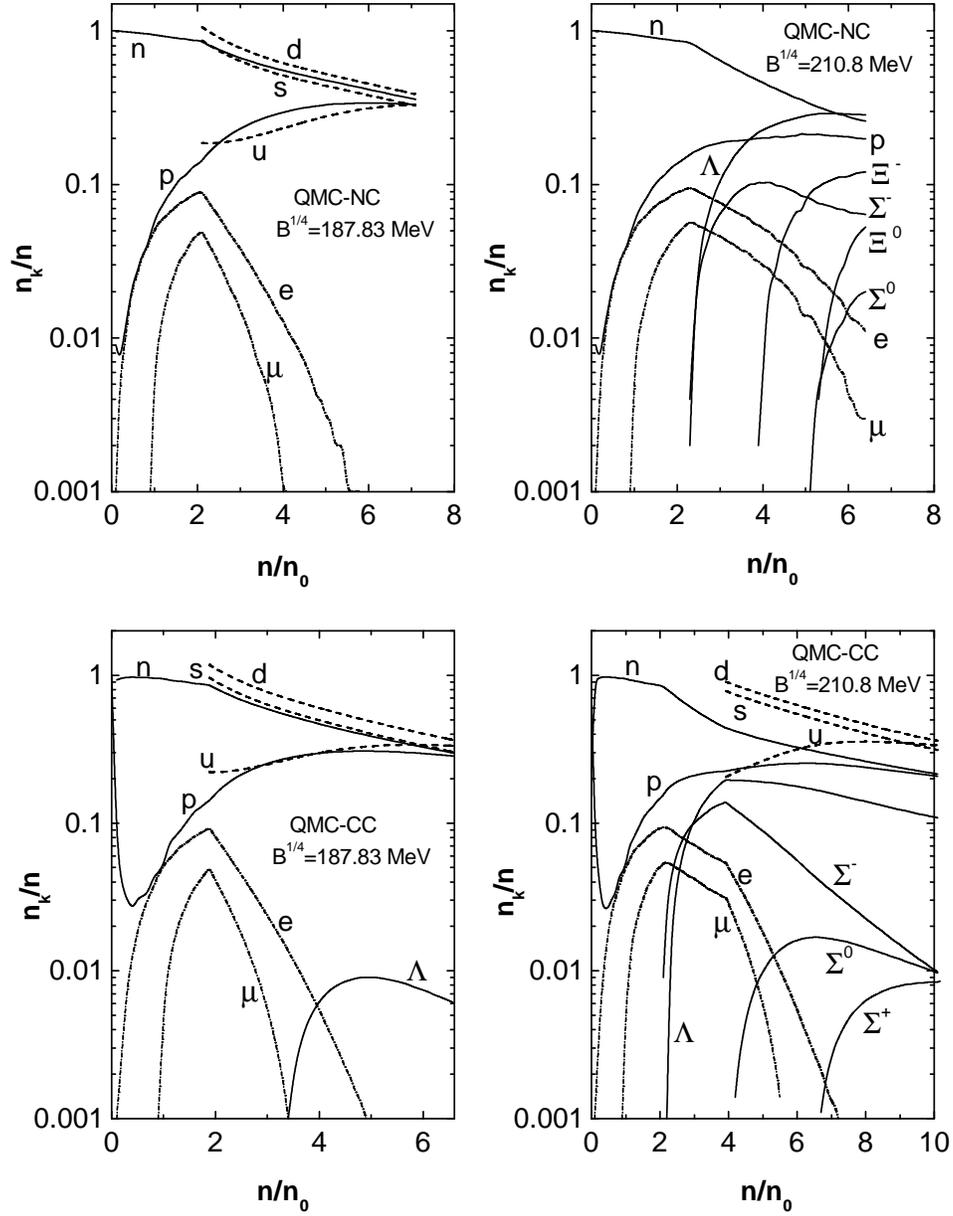}
 \caption{\footnotesize{The baryon and quark
composition of the star matter in terms of the baryonic density for
the bag constant values $B^{1/4}=187.8$ and $210.85$ MeV, for the
NC and CC cases. The line convention is explained in each
panel.}}\label{FIG4}
\end{figure}

 The composition of star matter is
depicted in Fig. \ref{FIG4} for cases (b) and (c). The onset of
the quark phase in the NC instance suppresses the hyperons that
could be present if this transition would not happened. Otherwise
in the CC approach new hyperon species appear even in the mixed
phase. It must be noted the pronounced decrease of the lepton
abundance in the mixed phase.

\begin{figure}[ht]
\centering
\includegraphics[width=0.8\textwidth]{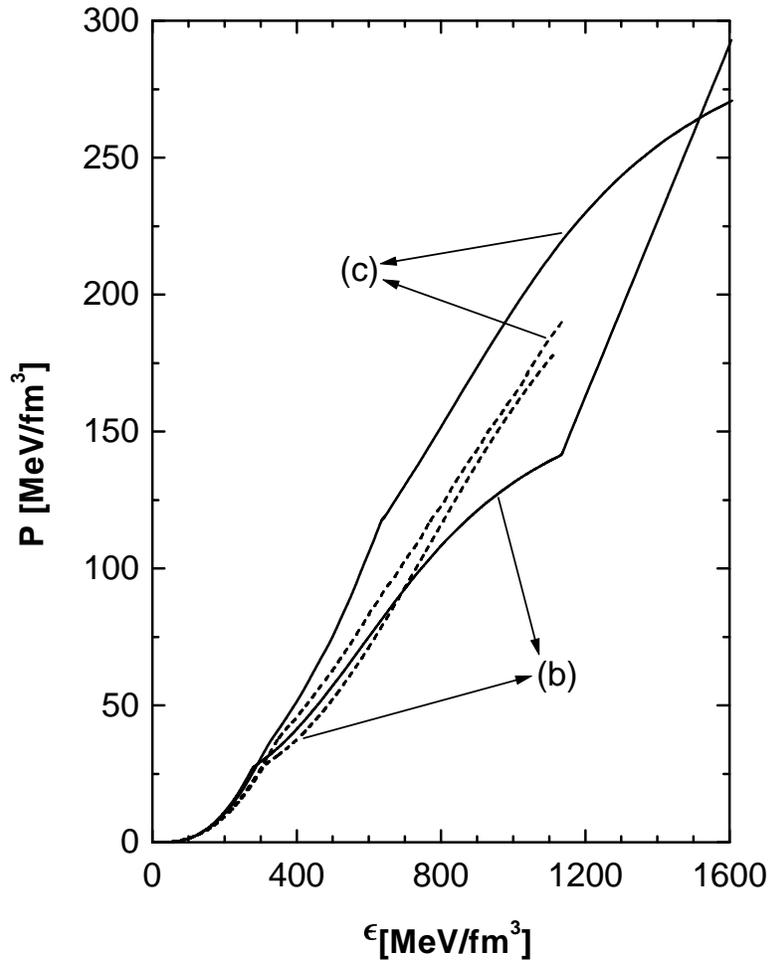}
 \caption{\footnotesize{The pressure $P$ of star matter in terms of
the total energy density $\epsilon$ = E/V, for the bag constant
values $B^{1/4}=187.8$ and $210.85$ MeV, for the NC (dashed lines)
and CC (solid lines) cases. For the last instance the pressure has
discontinuous derivatives at $n_{cl}$ and $n_{cu}$ (see Table
\ref{TABLEI}).}}\label{FIG5}
\end{figure}

The equation of state  for (b) and (c) is represented in Fig.
\ref{FIG5}. The energy density $\epsilon$ in the mixed phase
varies in the range $0.3 - 1.1 GeV fm^{-3}$ for (b)-CC, whereas it
ranges between $0.6 - 2. GeV fm^{-3}$ for the set (c)-CC. In the
CC treatment the pressure shows sudden changes of slope at the
extreme points of the mixed phase, which are absent in the NC
case.

\begin{figure}[ht]
\centering
\includegraphics[width=0.95\textwidth]{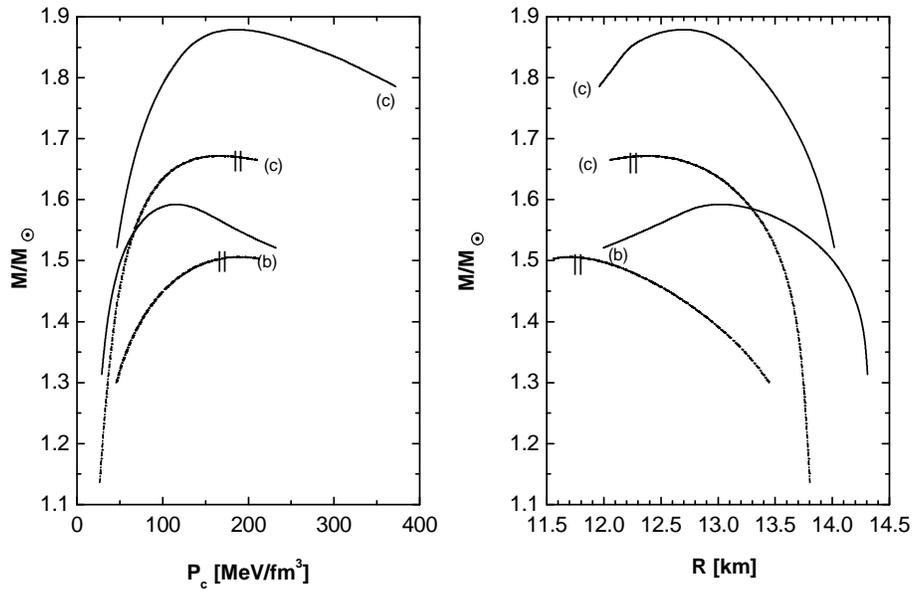}
 \caption{\footnotesize{The gravitational
star mass $M/M_\odot$ in terms of the central pressure $P_c$ (left
panel) and $M/M_\odot$ in terms of the star radius $R$ (right
panel). It can be appreciated that excluded volume corrections
enhance both $\overline{M}$ and $\overline{R}$, CC cases (solid
lines). Numerical values can be examined in Table~\ref{TABLEII}.
The double vertical bars have the same meaning as in Fig.
\ref{FIG2}, indicating that in the NC cases (dashed lines) the
bags at the center of the star are close to overlap for masses
near the maximum. }}\label{FIG6}
\end{figure}

The results for neutron star masses and radii based on these
equations of state are displayed in Fig. \ref{FIG6}. They are given
in units of the solar mass $M_\odot=1.9889 \, 10^{30} kg$.

The masses $\overline{M}$, radii $\overline{R}$ and moments of
inertia $\overline{I}$ for the maximum mass star are listed in
Table \ref{TABLEII}. We have found that $1.51\leq
\overline{M}/M_\odot\leq 1.88$, a result which is above the
experimental lower limit $M/M_\odot=1.44$ accepted for binary
pulsars.

For these particular cases we have obtained that the radii
$\overline{R}$ are rather insensitive to the star internal
structure, ranging between $12 \, km \lesssim \overline{R}
\lesssim 13 \, km$, being slightly bigger when excluded volume
corrections are included. Fig. \ref{FIG6} also shows that the
maxima are reached in a plateau, and they are enhanced in the CC
cases studied here with respect to the NC ones (Table
\ref{TABLEII}). It can also be appreciated what happens when
excluded volume effects are not considered. In fact, the NC EOS
(dashed lines) predicts that for stars approaching the maximum
stable mass $\overline{M}$ the baryonic bags at the center are
close to overlap. This is indicated by vertical double bars on the
corresponding curves. In particular for the case (b)-NC the bags
are overlapping at the center of $\overline{M}$, see Table
\ref{TABLEI} ($n_{max}$) and Table \ref{TABLEII} ($n_{c max}$). On
the other hand, when CC corrections are included no overlap is
present and this is one of the points in favor of the inclusion of
the excluded volume correlations.

At this point we can compare our predictions with previous works,
as for example \cite{PAL} where the QMC model is used to study
hyperonic matter and neutron star properties. There the bag
constant depends on the scalar mesons fields, and the increase of
the radius with density leads to an early overlap of the bags.
This feature contrasts with the monotonous decrease of the bag
radii we have found. We can also examine the EOS, although the
transition to quark matter is not considered in \cite{PAL}. The
model named QMCI in this reference and the set (b)-NC predict a
similar EOS, and for the option (b)-CC we find a slightly stiffer
behavior in the pure hadronic matter region due to excluded volume
corrections. At higher densities the appearance of the mixed phase
softens significantly the EOS, as it is shown in Fig. \ref{FIG5}.
With respect to the maximum neutron star mass and size in the case
(b)-NC (Table~\ref{TABLEII}) compares fairly well with those given
by the model QMCI \cite{PAL}.

\begin{figure}[ht]
\centering
\includegraphics[width=0.95\textwidth]{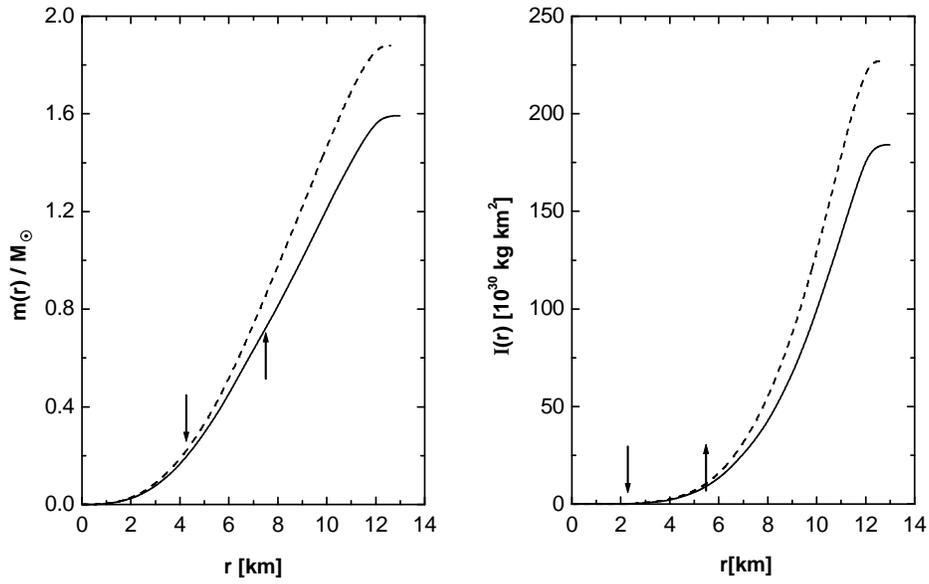}
\caption{\footnotesize{The star mass (moment of inertia) enclosed
within the spherical region of radius $r$ (see Eqs. (\ref{MASSR})
and (\ref{INERR})) is shown in the left (right) panel for the sets
(b) (solid line) and (c) (dashed line) within the CC treatment.
These results correspond to the maximum star mass $\overline{M}$,
as it is listed in Table~\ref{TABLEII}. The mixed phase threshold
is marked with a vertical arrow.}} \label{FIG7}
\end{figure}

The structure of the star, in terms of shells of either pure
quark, mixed quark-hadron, or pure hadron matter, depends on the
selected value of the bag parameter $B$. This can be seen in Fig.
\ref{FIG7} for the CC (b) and (c) cases, where we plot the mass
$m(r)$ (left panel) and moment of inertia $I(r)$ (right panel)
enclosed within a given $r$, for the maximum star masses
(Table~\ref{TABLEII}). The mass $m(r)$ is given in Eq.
\ref{MASSR}, and $I(r)$ corresponds to the definition given in Eq.
\ref{INERR} with the replacement $R \rightarrow r$ in the upper
integration limit. As indicated in this figure, the case (b)-CC
predicts that a mixed phase of deconfined quarks and hadrons can
be found in a central core with a radius of about $7.48km$, which
contains $45\%$ of $\overline{M}$ and contributes $18\%$ to the
moment of inertia $\overline{I}$ of the star. Meanwhile in the
(c)-CC case the hadron-quark mixed phase extends up to $4.31km$
and it encloses only $12\%$ of $\overline{M}$; its contribution to
$\overline{I}$ is small, around $1.5\%$.

For the sake of completeness, we mention that the option (a)-CC,
although not physically reliable, predicts for the star with the
maximum mass a structure which includes a pure quark core of
$6.55km$. This core is surrounded by a crust of mixed phase
located between $6.55km$ and $8.04km$, and the outer shell
contains only hadronic matter. The mass of this star is $1.42
M_\odot$ with a radius of $9.28 km$.

\section{Conclusions}

We have studied the high density regime of matter in a schematic
model of quarks confined into bags, the so called QMC model. One
of the main hypotheses of this model is that bags do not overlap,
but this situation is reached at relatively low densities of
around three times the nuclear matter saturation density in the
fixed $B$ treatment. Two different approaches have been given to
solve this problem by relaxing the requirement of non-overlapping
bags \cite{SR, PAL}. However these procedures could give rise to
formal inconsistencies such as the non-validity of the boundary
conditions at the bag surface. This point should be clarified
before practical applications at extreme densities can be done.

In the present work we intend to retain all the formal aspects of
the original QMC, including at the same time the dynamical effects
arising at very high densities. We do this by modifying the
standard QMC treatment in two aspects, in first place we consider
an alternative condition for the equilibrium of a bag immersed in
a dense medium. On the second place we introduce in-medium
averaged short range correlations among baryons by using an
excluded volume treatment. Thus we obtain a quasi-particle picture
of baryons dressed by these strong correlations, that remains
valid in all the range of the present calculations.

We studied the equation of state of hadronic matter and the
transition to a deconfined quark phase.  Our approach links
coherently the nonperturbative QCD effects represented by the bag
constant $B$, for both phases of confined and deconfined quarks.
The excluded volume correction becomes effective at intermediate
densities, contributing to the onset of a quark matter phase. The
non-overlapping bags hypotheses is verified over all the range of
densities for which hadrons are the relevant degrees of freedom.

We have found that for star matter a mixed hadron-quark phase
precedes the pure quark matter, in agreement with previous
studies.  It is also found sudden changes in the compressibility
at the extreme points of this mixed phase, as well as abrupt
changes in the density rate of growth of the vector iso-vector
meson and in the lepton concentration.

The effects of the finite baryon volume correlations on the
neutron star structure have been examined, we have found that the
maximum star mass is enhanced by both increasing $B$ and
introducing excluded volume corrections. The CC approach leads to
a small increase of the star radius at the maximum mass, as
compared to the respective NC case.

We have demonstrated that the CC treatment enlarges the range of
applicability of the QMC model, which can be extended to reach the
pure quark matter phase. In effect, in our approach this
transition takes place at baryonic densities where excluded volume
corrections are significative, and they should be included to
describe properly the structure of massive systems, like neutron
stars.

Our results are based on a schematic model, however it seems to
include the main ingredients for a reliable qualitative
description of the high density regime of matter. Further
refinements can be introduced to give more elaborate predictions,
such an improved treatment of the quark matter phase and the
consideration of exotic multiquark states, which will be object of
future studies.

\newpage
\begin{table}[ht]
\centering
\begin{tabular}{|l|c|c|c|c|c|c|c|c|} \hline
$B^{1/4} [MeV]$ & $R_p [fm]$& Case &  $g_{\sigma}^{u,d}$ &
$g_{\omega}^{u,d}$ & $g_{\rho}^{u,d}$ & $ n_{cl}/n_0$ &
$n_{cu}/n_0$ & $n_{max}/n_0$ \\ \hline 169.93& 0.8&
NC&5.747&2.756&8.668&1.20&--&3.40
\\ &&CC&4.678&1.701&7.600&1.06&3.98&--\\\hline
187.83&0.7&NC&5.858&2.872&8.582&2.10&--&6.40\\
&&CC&5.311&2.372&7.948&1.87&6.58&--\\\hline
210.85&0.6&NC&5.993&3.007&8.523&--&--&6.20\\
&&CC&5.702&2.747&8.151&3.91&10.17&--\\ \hline
\end{tabular}
 \caption{\footnotesize{The quark-meson couplings
$g_{\sigma, \omega, \rho}^{u,d}$ for each of the three values of
the bag constant $B$ used in our calculations, and for each of the
approaches with excluded volume correction (CC) and without it
(NC). The following two columns show the lower ($n_{cl}$) and
upper ($n_{cu}$) densities of the mixed hadron-quark phase, and in
the last column the limiting density $n_{max}$ for the NC cases
are displayed.}}\label{TABLEI}
\end{table}

\vspace{5cm}
\begin{table}[ht]
\centering
\begin{tabular}{|l|c|c|c|c|c|c|} \hline
$B^{1/4} [MeV]$ & Case  &$\overline{M}/M_\odot$ & $\overline{R} [km]$&
$n_{c \, max}/n_0$ & $\overline{z}$ & $\overline{I}[10^{30} kg \, km^2]$\\
\hline
187.83&NC&1.506&11.70&6.886&0.2701&137.8\\
&CC&1.592&13.02&5.148&0.2513&184.1\\ \hline
210.85&NC&1.672&12.38&5.800&0.2897&181.5\\
&CC&1.879&12.69&5.507&0.3330&227.1\\ \hline
\end{tabular}
 \caption{\footnotesize{Neutron star properties for
each of the bag constants used in our calculations and for
each of the approaches: with excluded volume correction (CC) and
without it (NC). The star mass $\overline{M}$ relative to the sun
mass, the star radius $\overline{R}$, the central baryonic density
($n_{c \, max}$), the surface redshift $\overline{z}$ and the
moment of inertia $\overline{I}$ (for a slowly rotating star), all
corresponding to the star with maximum mass. }}\label{TABLEII}
\end{table}

\newpage
\newpage

\clearpage

\newpage

\newpage

\newpage

\newpage

\end{document}